\documentstyle [12pt]{article}
\newcommand{\mincir}{\raise -2.truept\hbox{\rlap{\hbox{$\sim$}}\raise5.truept
\hbox{$<$}\ }}
\newcommand{\magcir}{\raise -2.truept\hbox{\rlap{\hbox{$\sim$}}\raise5.truept
\hbox{$>$}\ }}
\newcommand{\minmag}{\raise-2.truept\hbox{\rlap{\hbox{$<$}}\raise 6.truept\hbox
{$>$}\ }}
\newcommand{\be}{\begin{equation}}
\newcommand{\ee}{\end{equation}}
\newcommand{\ba}{\begin{eqnarray}}
\newcommand{\ea}{\end{eqnarray}}
\newcommand{\brr}{\begin{array}}
\newcommand{\err}{\end{array}}
\newcommand{\bc}{\begin{center}}
\newcommand{\ec}{\end{center}}

\newcommand{\bx}{\mbox{\bf x}}

\newcommand{\hm}{\,h^{-1}{\rm Mpc}}

\title{ {\bf Is there any Scaling in the Cluster Distribution ?}}
\author{
{\bf Stefano Borgani,}$^{1,2}$
{\bf Vicent J. Mart\'{\i}nez,}$^{3,4}$ \\
{\bf Miguel A. P\'erez}$^5$ \&
{\bf Riccardo Valdarnini}$^2$ \\ ~\\
{\em $^1$ INFN -- Sezione di Perugia, c/o Dipartimento di Fisica
dell'Universit\`a} \\ {\em Via A. Pascoli, I--06100 Perugia, Italy} \\ ~\\
{\em $^2$ SISSA -- International School for Advanced Studies,} \\ {\em Via
Beirut 2--4, I--34014 Trieste, Italy}\\ ~\\
{\em $^3$ Departament de Matem\`atica Aplicada i Astronomia,} \\ {\em
Universitat de Val\`encia,} \\ {\em E--46100 Burjassot, Val\`encia,
Spain} \\ ~\\
{\em $^4$ Astronomy Unit, School of Mathematical Sciences,} \\ {\em Queen
Mary \& Westfield College,} \\ {\em Mile End Road, London E1 4NS, UK} \\ ~\\
{\em $^5$ Sternberg Institute of Astronomy, Moscow State University,} \\
{\em Moscow, Russia} }

\date{}

\begin{document}
\maketitle
\vspace{0.5truecm}
\centerline{\sl ApJ, accepted}

\vspace{1.truecm}
\section*{\center Abstract}
We apply fractal analysis methods to investigate the scaling properties in
the Abell and ACO catalogs of rich galaxy clusters. The methods are adapted
to account for the incompleteness of the samples by appropriately using the
known selection functions in the partition sums. We also discuss different
technical aspects of the method when applied to data sets with small number
of points as the cluster catalogs. Results are compared with simulations
based on the Zel'dovich approximation.

We limit our analysis to scales less
than 100 $\hm$. The cluster distribution show a scale invariant
multifractal behavior in a limited scale range. For the Abell catalog this
range is 15--60$\hm$, while for the ACO sample it extends to smaller scales.
Despite this difference in the extension of the scale--range where
scale--invariant clustering takes place, both samples are characterized by
remarkably similar multifractal spectra in the corresponding scaling regime.
In particular, the correlation dimension turns out to be $D_2\simeq 2.2$
for both Abell and ACO clusters.

Although it is
difficult to point out the scale at which homogeneity is reached with the
present size of these redshift surveys, our results indicate that the
cluster distribution shows a tendency to homogeneity at large scales,
disproving the picture of a pure scale invariant fractal structure
extending to arbitrarily large distances.

\vspace{0.3truecm}
\noindent
{\em Subject headings: Galaxies: clusters; large--scale structure}

\newpage
\section{Introduction}
The study of the large scale matter distribution is one of the fundamental
problems in modern cosmology. Extended redshift surveys of galaxies and
galaxy clusters have reveled the existence  of filaments and
sheet--like structures up to the maximum considered depths $\magcir
100\hm$\footnote{$h$ is the Hubble constant in units of 100
Km sec$^{-1}$ Mpc$^{-1}$} (Kirshner et al. 1981; de Lapparent et al. 1986;
Geller \& Huchra 1989; Broadhurst et al. 1990; Postman, Huchra \& Geller
1992). In the standard framework of the galaxy formation  theory these
patterns arise as a result of the gravitational interactions  in the
expanding medium, with primordial fluctuations of very low amplitude acting
as seeds for the subsequent structure formation.

It is then clear that the study of large--scale distribution of cosmic
structures is of extreme interest since it should give us information about
the shape and statistics of the primeval fluctuation spectrum. Catalogs of
different extragalactic objects  have been studied with various statistical
methods. One of the most popular approach is based on the estimate of the
correlation functions for both galaxy and cluster distributions. In this
respect, the most striking result is that the 2--point correlation
function, $\xi(r)$, turns out to closely follow a power--law shape,
\be
\xi(r)~=~\left({r_o \over r}\right)^{\gamma}\,,
\label{eq:xir}
\ee
with $\gamma\simeq 1.8$ for both galaxies and clusters, although holding on
different scale ranges and with remarkably different correlation length,
$r_o$; for galaxies it is $r_{o,g}\simeq 5$--6$\hm$ (Peebles 1980; Davis \&
Peebles 1983; Vogeley et al. 1992) for $0.1\mincir r\mincir 10\hm$, while
it is $r_{o,c}\simeq 20\hm$ for rich Abell clusters in the scale--range
$10\mincir r\mincir 60\hm$ (Bahcall \& Soneira 1983, Klypin \& Kopylov 1983;
Postman, Huchra \& Geller 1992). Although the exact value of the cluster
correlation length still represents a widely debated issue, with several
authors claiming a smaller value, $r_{o,c}\simeq 14\hm$ (Sutherland 1988,
Sutherland \& Efstathiou 1991, Dalton et al. 1992), nevertheless it is
firmly established that clusters are much more strongly correlated than
galaxies and the amount of this enhanced clustering represents a challenge
for any theoretical model of large--scale structure formation (e.g., White
et al. 1987; Bahcall \& Cen 1987; Holtzman \& Primack 1993; Mann, Heavens \&
Peacock 1993; Croft \& Efstathiou 1994; Borgani, Coles \& Moscardini 1994).
In the framework of the biased model for structure formation
(e.g., Kaiser 1984; Bardeen et al. 1986) the enhanced clustering of rich
galaxy systems is nothing but the consequence of their occurrence in
correspondence of high--density peaks of the underlying matter distribution.

By adopting a completely different point of view, Coleman, Pietronero \&
Sanders (1988) argued that the existing relation between the correlation
properties of galaxies and clusters is the probe for a self--similar
(fractal) structure of the Universe, extending up to arbitrarily large
scales. The corresponding fractal dimension, $D$, turns out to be related
to the slope of the 2--point correlation function according to $D=3-\gamma$.
In this picture, the absence of any characteristic scale,
beyond which homogeneity is reached, makes the correlation length $r_o$ a
meaningless quantity, since it turns out not to deal with intrinsic
properties of the clustering, but depends only on the size $R_s$ of the
sampled volume, according to $r_o\propto R_s$ (see Coleman \& Pietronero
1992, for a review about a fractal Universe). Therefore, this should
explain the depth dependence of the galaxy clustering (Einasto, Klypin \&
Saar 1986; Davis et al. 1988; Mart\'{\i}nez et al. 1993),
as well as the large value of the cluster
correlation length, as due to the larger volumes encompassed by cluster
samples with respect to galaxy samples.

As for the galaxy distribution, Mart\'{\i}nez \& Jones (1990) have shown that
any self--similarity should be confined to small scales, $r_o\mincir 4\hm$
(see also Mart\'{\i}nez et al. 1990), while homogeneity should be reached at
larger scales. It is not clear whether or not such scales are encompassed by
presently available redshift samples. The view of a small--scale fractality
followed by a smooth transition toward large--scale homogeneity has also
been supported by scaling analyses of the non--linear clustering developed
by N--body simulations (Valdarnini, Borgani \& Provenzale 1992; Colombi,
Bouchet \& Schaeffer 1992; Yepes, Dom\'{\i}nguez--Tenreiro \& Couchman 1992;
Murante et al. 1994, in preparation).
These analyses consistently show that at small scales of few Mpcs
non--linear gravitational dynamics acts to create a self--similar
clustering. The resulting fractal dimension turns out to take a
characteristic value, $D\simeq 1$, quite independent of the choice for the
initial fluctuation spectrum. Therefore, the detected value of the galaxy
correlation length must not be interpreted as the scale above which
homogeneity is attained. Instead, it separates the small--scale regime of
non--linear gravitational clustering from the larger--scale one, where
mildly non--linear dynamics preserves a more clear imprint of initial
conditions.

Due to the similarity between the clustering properties of galaxies and
clusters, one may ask whether the observed self--similarity for $r\mincir
r_o$ also holds for the cluster distribution. If this were the case, it is
clear that it cannot be explained on the ground of non--linear gravitational
clustering, since it occurs at scales ($\magcir 10\hm$) where the
fluctuation evolution should still be near to the linear regime.

Actually, Borgani, Plionis \& Valdarnini (1993a) analyzed the scaling
properties for the projected distributions of cluster samples having
different richness, selected with an overdensity criterion from the Lick
galaxy map (Plionis, Barrow \& Frenk 1991). As a result they found that the
distribution of the richest clusters displays a well defined scaling
behavior, remarkably similar to that of galaxies, for angular scales
$\vartheta \mincir 7^\circ$ (corresponding to physical scales $\mincir
25\hm$ at the depth of the Lick map). At larger scales, the
self--similarity breaks down and homogeneity, at least in the projected
distribution, is attained. Furthermore, it has also been shown that
simulating a cluster distribution by projecting a three--dimensional
scale--invariant structure does not introduce characteristic scales in the
angular distribution, thus indicating that the breaking of self--similarity
in the angular distribution reflects the presence of a characteristic scale
even in the spatial cluster distribution.
Borgani et al. (1994) generated synthetic angular samples by projecting
three--dimensional N--body simulations based on both Gaussian and skewed
initial conditions for a CDM spectrum. After reproducing the observational
setup of the Plionis et al. (1991) cluster samples, they addressed the
question whether the detected small scale self--similarity for clusters
imposes a stringent constraint on the initial conditions. As a result they
found that, at least for the CDM model, the initial Gaussian statistics
does not succeed at explaining the small--scale fractality, while some
non--Gaussian models fare much better.

In the present paper we realize a detailed scaling analysis for large
redshift samples of Abell and ACO clusters and will compare them with
synthetic cluster samples extracted from numerical simulations based on
Zel'dovich simulations with Gaussian initial conditions. This kind of
analysis should allow us to answer the following questions: $a)$ is the
scale--invariance detected for angular samples at small scales a real
feature imprinted in the three-dimensional distribution ?; $b)$ If yes,
can it be explained simply on the ground of a peak selection procedure on a
Gaussian background or does it require something more ?; $c)$ Do
the present cluster samples encompass the scale of homogeneity or should
the fractal picture of the Universe be not still disproved ?

The plan of the paper is as follows. In section 2 we present the real
cluster samples and the simulations of the cluster distributions.
In section 3 we study the scaling in the high density regions by
means of the correlation integral method for real catalogs as
well as simulations. Section 4 is devoted to the analysis of the
scaling in the low density regions by means of the $n$--nearest
neighbor distances. General conclusions are drawn in section 5.

\section{The cluster samples}

\subsection{Real data}
The analysis is done on subsets of the Abell and ACO cluster catalogs
(Abell 1958; Abell, Corwin and Olowin 1989). The subsample parameters are
those already used in previous papers (Plionis \& Valdarnini
1991). For the Abell catalog the geometrical boundaries are
$|b^{II}|\ge 30^{\circ}$,  $\delta \leq -27^{\circ}$ and  $z \leq 0.1$,
with a total area of 4.8sr. These constraints result in a total number of
206 clusters distributed in the Northern (NGC) and Southern (SGC) Galactic
Cap. For the ACO catalog the selection criteria are $m_{10} \le 16.4$,
$b^{II} \le -20^{\circ}$ and $\delta \geq -17^{\circ}$.
The subsample has an area of 1.8 sr and 103 clusters. In this subset 19
clusters have redshift estimated form the $m_{10}-z$ relation used by
Plionis \& Valdarnini 1991, otherwise redshift are directly taken from the
original catalog or complemented with those measured by a number of authors
(Plionis, Valdarnini \& Jing 1992, hereafter PVJ). In the above selection
criteria we included all cluster with richness $R \geq 0$. According to
Postman, Huchra \& Geller (1992) the density distribution and the
clustering properties of the $R=0$ and $R\ge 1$ cluster samples are similar
out to $z\approx 0.2$. In a different context, PVJ have checked the
robustness of their statistical analysis by applying the same tests to the
$R \geq 1$ and $R \geq 0$ sample. The results are not very much different
and we decided to include clusters with $R=0$ in our studies. The selection
criteria we used allow us to have defined subsets of the original cluster
catalogs almost volume--limited, excluding projection effects due to poorly
sampled, rich and distant clusters.

The selection function we introduce are both in galactic latitude and
redshifts. The former has the functional form
\be
P(b^{II})~=~10^{\alpha(1-csc(|b^{II}|))},
\label{eq:pb}
\ee
where $\alpha=0.3$ ($0.2$) for the Abell (ACO)
subsample. As for the redshift selection function, we take the expression
\be
P(z)=\cases{1&$~~~;~~~z<z_c$\cr
		A\exp^{-z/z_0},&$~~~;~~~z\geq z_c$\cr}\,.
\label{eq:pz}
\ee
Here $z_c$ is the maximum redshift at which the cluster distribution
follows that of a uniform one. The choice of $z_c$ is given by a
lest--square fit of $P(z>z_c)$ vs. $\log{N_{obs}/N_{exp}}$, where $N_{obs}$
and $N_{exp}$ are the number of clusters which are observed and which are
expected for a uniform distribution, respectively. The redshift are
$z_c=0.081, 0.063$ and $0.066$ for NGC, SGC and ACO, respectively. The
details of the selection function determination are described in PVJ.

In the following, the Abell North and Abell South samples will be merged
together, while the ACO clusters will be considered separately, since they
have been shown to display  some different clustering properties (PVJ;
Cappi \& Maurogordato 1992).

\subsection{Simulations}
The simulated cluster catalogues that we use are the same of PVJ. We refer
to this paper for more details. We randomly distribute $N_p$ points in a
cube of size 640 $h^{-1}$ Mpc with $N_g^3=64^3$ grid points. The points are
then displaced from their positions  according to the Zel'dovich
approximation (Zel'dovich 1970; Shandarin \& Zel'dovich 1989).
The Zel'dovich approximation has been shown
to be an adequate prescription to follow gravitational clustering in the
mildly non--linear regime (e.g., Coles, Melott \& Shandarin 1993). Its
relevance to the study of the distribution of galaxy clusters has been
stressed by Blumenthal, Dekel \& Primack (1988) and, more recently, by
Mann et al. (1993) and Borgani et al. (1994).

The density fluctuation spectrum $\delta_{\vec k}$  is chosen to have
a Gaussian distribution with random phases and power spectrum
\be
P(k)~=~\langle \delta_{\vec{k}}^2 \rangle~=~A\,k^n\,\exp{(-|{\bf k}|^2/
\Lambda^2)}\,\Theta(|{\bf k}|)\,.
\label{eq:pk}
\ee
Following Postman et al. (1989), we take $\lambda^{-1}=0.1\hm$, $\Theta(
|{\bf k}|) =0$ for $|{\bf k}|>(80\hm)^{-1}$ and $\Theta(|{\bf k}|)=1$
otherwise.

To each particle is tagged a $\nu$ value, such that $\nu =\delta_{\vec g}/
\sigma$, being $\delta_{\vec g}$ the value at the nearest grid point of the
fluctuation of the linear density field smoothed with a Gaussian window of
size $R=10\hm$, and $\sigma$ the corresponding rms fluctuation value.

The number of points in the simulations, $N_p$, must be chosen so that,
after having applied to the simulated samples the survey boundary limits
and all the selection functions of the real data, we have the same number
of clusters as in the real samples. Taking $N_p = 73000$ results to about
$\sim 12000$ points being associated with peaks with $\delta > \nu \sigma$
with $\nu =1.3$.

The parameters of the simulations, namely $A$, $n$ and $\nu$, are chosen in
such a way to reproduce the observed cluster 2--point correlation function
in the appropriate scale range. PVJ generated several cluster simulations,
all having the same slope $\gamma \simeq 1.8$ for $\xi(r)$, and clustering
lengths ranging from $\simeq 14\hm$ to $\simeq 25\hm$. In the present
analysis we will only consider the simulation with $r_o\simeq 20\hm$, which
has been shown to better reproduce several clustering features of the Abell
and ACO samples. A set of 50 different initial phase assignments is
considered, so to have a large ensemble over which to
estimate the effects of the {\em cosmic variance} for the cluster
distribution. PVJ have shown that these cluster simulations, which are
called to reproduce the observed 2--point function, also reproduces as an
extra bonus even the 3--point function, thus supporting their
reliability to give a faithful representation of the clustering of rich
galaxy clusters.

\section{Scaling from the correlation integral analysis}
\subsection{The method}
Fractal analysis methods are based on the determination of the so--called
spectrum of generalized dimension, which characterizes the scaling
properties of the system. In order to introduce the concept of fractal
structure and fractal dimension, let us consider a point distribution and
suppose to cover it with cubic cells all having the same size $r$. Let also
$\bar p_j(r)$ be the probability measure associated to the $j$--th cell, that
is
the fraction of all the points inside there, and $N_c(r)$ the number of
non--empty cells of side $r$. Accordingly, for a fractal structure it is
possible to define the sequence of R\'enyi dimensions
\be
D_q~=~{1\over q-1}\,\lim_{r\to 0}{\log\,\sum_{j=1}^{N_c(r)}[\bar
p_j(r)]^q \over \log\,(1/r)}
\label{eq:re}
\ee
(R\'enyi 1970), $q$ being a generic real number. For $q=0$, eq.(\ref{eq:re})
provides the box--counting (capacity) dimension, which only depends on the
number of non--empty cells, thus accounting only for the geometry of the
distribution. Note that for $q>0$ the sum in eq.(\ref{eq:re}) mostly
weights the overdensities, while $q<0$ is for the underdense parts of the
distribution. Under general conditions it is possible to show that $D_q$ is
a non--increasing function of $q$. The simplest case occurs when
$D_q=const$. In this case the structure is said to be monofractal and it is
described just by a single scaling index (see Paladin \& Vulpiani 1987, for
a review about multifractals).

The interpretation of large--scale clustering in terms of fractal dimensions
has been shown to provide a comprehensive description, the $D_q$
dimension spectrum of eq.\ref{eq:re}) being strictly connected to
fundamental statistical descriptors, such as the $N$--point correlation
functions and the void probability function (e.g., Balian \& Schaeffer
1989, Jones, Coles \& Mart\'{\i}nez 1992; Borgani 1993).

Although the definition (\ref{eq:re}) of $D_q$ dimension spectrum is given
in the limit of infinitesimally small scales, in practical estimates one
usually deals with a finite amount of data points, so that only finite
scale--ranges can be probed. Therefore, one is forced to resort to
approximate algorithms, which are based on different approximations to the
formal definition of fractal dimensions and suffers by different degrees
for the presence of poor statistics (see, {\em e.g.}, Mart\'{\i}nez et al.
1990; Valdarnini, Borgani \& Provenzale 1992; Borgani et al. 1993b). One of
such methods is based on the evaluation of the correlation integrals
(Grassberger \& Procaccia 1983; Paladin \& Vulpiani 1984): for each point
$i$, one evaluates the probability measure $p_i(r)$, which represents the
fraction of all the points within the sphere of radius $r$ and centered on
$i$. Accordingly, the statistics is described by the partition function
\be
Z(q,r)~=~{1\over N}\,\sum_{i=1}^N[p_i(r)]^{q-1}~=~
{\left<n^{q-1}\right>\over N^{q-1}}~\sim ~r^{\tau_q}\,.
\label{eq:gp}
\ee
Here $N$ is the total number of points in the sample and $\left<
n^q\right>$ is the $q$--th order moment for the neighbor counts. For a
fractal structure the scaling index $\tau_q$ is strictly independent of the
scale and defines the $q$--th order dimension as
\be
D_q~=~{\tau_q\over q-1}\,.
\label{eq:dq}
\ee
Although this method is rather reliable to estimate fractal dimensions when
a high number of points is available, spurious estimates are originated by
limited statistics. This problem is expected to be particularly important
for the distribution of clusters, whose mean separation ($d\simeq 40\hm$) is
roughly twice the corresponding correlation length ($r_o\simeq 20\hm$),
but not as important for the galaxy distribution, for which $d\simeq r_o
\simeq 5\hm$.

A subtle point concerning the evaluation of the $p_i(r)$ probabilities is
whether or not including in it the center point $i$. It is clear that, when
the number of neighbors is sufficiently large, no difference is expected,
while  much more care is needed at the smallest sampled scales. Grassberger
(1988)  strongly recommended that that "the central point must obviously
not  included". In this way, for $q>1$ the sum in eq.(\ref{eq:gp}) always
weights  only the more clustered structures, thus limiting the bias in the
dimension  estimate due to the poorly sampled part of the distribution.
However, as smaller and smaller scales are considered, more and more points
give no  contribution to $Z(q,r)$. Consequently, the partition function
steepen at  small scales and the resulting dimension diverges.

On the other hand, by including the center point in the estimate of
$p_i(r)$,  at very small scales most spheres will take only the
contribution from the  center. As a consequence, the value of the partition
function changes only  slowly by increasing $r$, so that $D_q\simeq 0$.
Although spurious,  in this case this result has a simple geometrical
interpretation, since at small  scales we are measuring nothing but the
topological dimension of each single  point, which is indeed zero.

An efficient method to correct dimension estimates from undersampling lies
in subtracting the shot--noise contributions, which generates the
underestimate of the local dimension. Let us denote by $d\mu(x)$ the
local measure associated to the fractal structure. Therefore, the ``mass"
contained within a ball of radius $r$, centered on the point $i$ belonging
to the fractal, is $\mu_i(r)=\int_{<r}d\mu$. If
$m_q(r)=\left<\mu^q(r)\right>$ is the corresponding $q$--th order moment,
then, under the usual assumption that this sampling is purely Poissonian,
the moments of counts for the sampling points, $\left< n^q\right>$, are
related to the ``true" moments, $m_q$, according to suitable recurrence
relations. At the first orders these read
\ba
\left< n^2\right> & = & \left<n\right> + m_2\,; \nonumber \\
\left< n^3\right> & = & \left<n\right> + 3m_2 + m_3\,; \nonumber \\
\left< n^4\right> & = & \left<n\right> +7m_2 +6m_3 +m_4
\label{eq:pois}
\ea
({\em e.g.}, Peebles 1980), while more cumbersome expressions are expected
at higher orders. The recursive application of the above expressions allows
one recover the background statistics from that of the Poissonian
realizations. The reliability of this prescription in fractal analysis to
recover the correct dimension estimates has been checked by Borgani \&
Murante (1994) and shows that the expected scaling is
nicely recovered even when heavily diluting the original sample. It is
however clear that such a procedure to correct for poor sampling does not
always guarantee a meaningful result. Typical examples are when the mean
neighbor count is much less than unity or when the points are located in
correspondence of density peaks, which clearly do not represent a Poissonian
sampling.

A further problem in the application of the correlation--integral method
resides in the treatment of boundaries, when the point distribution is
confined in a finite volume. A first possibility, which makes no
assumptions about the distribution outside the sample volume, consists in
discarding from the partition sum of eq.(\ref{eq:gp}) at the scale $r$ all
the centers whose distance from the nearest boundary is less than $r$.
However, this kind of procedure severely limits the
statistics and, moreover, the analyzed sample is statistically different at
different scales. This represents a serious problem in our case, since the
geometry of the boundaries and the limited number of clusters does not
allow to consider scales larger than $\sim 70\hm$. A further possibility is
to consider for each center $i$ the fraction $f_i(r)$ of the sphere of
radius $r$ centered on it, which falls within the boundaries. In this way,
if $\tilde n_i(r)$ is the counted number of neighbors, the corrected
values is $n_i(r)=\tilde n_i(r)/f_i(r)$. Actually, this procedure also
allows to account for other systematics of the cluster sample, such as the
dependence of the local cluster density on redshift and galactic latitude,
as provided by the selection functions of eqs.(\ref{eq:pb}) and
(\ref{eq:pz}).

However, this border correction method relies on the assumption that the
behaviour of the cluster distribution inside the sample boundaries is
statistically equivalent to that inside such boundaries. Therefore, it
assumes, rather than verifies, that the analyzed distribution represents a
fair sample. Therefore, one must be sure that, at the considered scales, no
serious effects of homogeneization are spuriously introduced by boundary
corrections.

The corrected local count around the $i$--th point is
\be
n_i(r)~=~{1\over f_i(r)}\,\sum_{j=1}^N{\theta (|\bx_i-\bx_j|-r)
\over P_j(b)\,P_j(z)}\,,
\label{eq:ncor}
\ee
where $P_j(b)$ and $P_j(z)$ are the values of the galactic latitude and
redshift selection functions at the position of the $j$--th cluster. In
order to implement this correction, we realized a Montecarlo sampling with
rejection, according to sample boundaries and selection functions as
described in section 2, so to measure the corrected count $n_i(r)$. Note,
however, that at distances much larger than the completeness redshift,
$P(z)$ rapidly declines, with a subsequent increase of the noise in the
correction procedure. For this reason, we will mainly present results based
on clusters within the distance $d=200\hm$, although we will also show the
effect of correcting for selection functions, when their value becomes much
smaller than unity.

In order to check that our analysis method correctly detects the presence
of a characteristic scale in the distribution, we resort to the
$\beta$--model prescription ({\em e.g.}, Paladin \& Vulpiani 1987) to
generate a scale--dependent monofractal structure with $D=1$ below some
homogeneity scale $L_h$ and $D=3$ above it. We choose a box--counting
homogeneity scale $L_h=80\hm$, which roughly corresponds to the homogeneity
scale $\simeq 40\hm$ as seen by the correlation--integral algorithm.
Therefore, we introduce the same boundaries and selection effects as for
real samples, so to generate synthetic data sets. In Figure 1 we plot the
local dimension $D_q(r)$, obtained from a five point log--log local linear
regression on the partition function shape, for the simulated Abell North
sample. From top to bottom we show results for progressively increasing $q$
values ($q=2,3$ and 5). Circles and squares correspond to including and
excluding the center object, respectively, in the estimate of the
probability $p_i(r)$. Triangles refer to the shot--noise corrections, that,
according to eq.(\ref{eq:pois}), can be applied for $q\ge 3$. For $q=2$,
including the center increases the shot noise effects associated to the
discrete nature of the distribution; $D_2$ is heavily underestimated at
small scales and no scale range is detected where the expected fractal
scaling takes place. Going to higher $q$ values, the method based on
including the center improves to some extent, although it still provides a
biased estimate of the local dimension. However, the reliability of this
method is largely improved after the removal of the shot--noise terms (see
also Borgani \& Murante 1994); in
this case the scaling regime is fairly detected in the expected range and
the correct fractal dimension value is recovered. This shows the
reliability of the method to detect the presence of a fractal scaling
regime. In a similar way, also excluding the center allows a fair
determination of the scaling regime.

Although the presence of the characteristic scale at $L_h\simeq 40\hm$ is
well detected, nevertheless even at the largest sampled scale the
local dimension does not attain the homogeneity value $D_q=3$. Instead, at
$r\simeq 90\hm$ it is $D_q\simeq 2$, quite independent of the
multifractal order $q$ and on the analysis method. This indicates some
inertia of the $Z(q,r)$ partition function to follow the change of scaling
when applied to a point distribution as large as the cluster samples we are
dealing with.

Furthermore, we verify that the boundary corrections do not introduce
characteristic scales in a structure which is otherwise self--similar at all
the scales. To do this, we generate a synthetic Abell North sample from a
$D=1$ monofractal structure without large--scale homogeneity. In Figure 2 we
plot the resulting local dimensions for $q=2,3$ and 5. In this case, only
results based on excluding the center point are presented, although no
significant differences between the methods have been detected, due to the
strong clustering associated to this pure fractal. It is apparent that
quite large fluctuations appear due to the lacunarity ({\em e.g.}, the
presence of big voids) induced by the $\beta$--model. However, no
homogenization is spuriously induced, at least at the scales of interest,
which, instead, would turn into $D_q\simeq 3$ at the largest considered
scales.

Based on the analyses presented in this section about the reliability of the
implementation of the correlation--integral method, we draw the following
conclusions.
\begin{description}
\item[a)] Excluding the center points in the estimate of $p_i(r)$ does usually
provide a fair dimension estimate. However, as we will see in the
following, a spurious increase of $D_q(r)$ should be expected in some cases
at the small scales, where undersampling becomes severe.
\item[b)] Including the center points could heavily underestimate the
dimension in a broad scale range and pollute the presence of a scaling
regime. However, after suitably removing the shot--noise terms, a reliable
estimate of the fractal dimension is provided and the scaling is correctly
recovered.
\item[c)] Our prescription to correct boundary effects is shown to fairly
detect the presence of a characteristic scale. However, the local dimension
fails to detect homogeneity at the largest scales sampled by the available
cluster samples, even for a structure which is known {\em a priori} to be
homogeneous at such scales. This shows that the condition $D_q\simeq 3$ is
sufficient but not necessary to claim homogeneity in the cluster
distribution.
\end{description}
Based on these results, in the following we will discuss the
results of our fractal analysis on real as well as simulated cluster
samples.

\subsection{The results}
In Figures 3 and 4 we plot the local dimension for different $q$ values,
for Abell and ACO clusters, respectively. Each panel is for a
given $q$ value and report the results for the partition function and for
the corresponding local dimension, as obtained by excluding the
contribution of the $i$--th cluster in the estimate of $p_i(r)$. For $q=3$
and 5 we also show the local dimension as obtained after correcting for
shot--noise effects, according to eq.(\ref{eq:pois}).
The errorbars in the partition function represents the 1$\sigma$ scatter
between 100 bootstrap resamplings. Uncertainties in the local dimension are
three times the standard deviations in the 5--point weighted least--square
fitting in the partition function slope. For reason of clarity we do not
plot errorbars for shot--noise correction results, which are always
comparable to the plotted ones.

For the Abell clusters, both methods consistently show that the profile of
the local dimension does not remain flat over the whole considered
scale--range, thus disproving the picture of a purely fractal structure of
the large--scale clustering. At small scales, $D_q(r)$ suffers from the
limited statistics and decreases toward small values ($D_q(r)\mincir 1$),
due to the flattening of $Z(q,r)$ at $r\mincir 10\hm$. On the other hand,
the dimension increases at $r\magcir 50\hm$ and reaches $D_q\magcir 2$ at
the largest considered scales ($\simeq 100\hm$). In Table 1, we show the
dimension values $D_q$ for $q=2,3$ and 5 calculated in the range
15--60$\hm$, by excluding the center point in the $n_i(r)$ local count.
Also reported are the 1$\sigma$ uncertainties.

Quite different results are found for the ACO clusters, for which both
methods do not indicate a smooth increasing of $D_q(r)$. For $q=2$,
excluding the centers in the $p_i$ estimates causes a spurious increase of
the $D_q$ at small scales. For higher $q$'s, more dense regions are
weighted and undersampling effects are less important. Correspondingly, the
same method detect the presence of a wider scale range, where the local
dimension remains nearly flat. Note that for $q=3$ correcting for
shot--noise reveals the existence of scaling regime, extending down to the
smallest considered scales, thus much wider than that detected for the
Abell clusters. Correspondingly, the $Z(q,r)$ partition function never
flattens even at $r<10\hm$. This result calls for a substantial difference
between the clustering properties of the two samples.
The self--similarity is broken at scales
$\simeq 50\hm$ after which the dimension starts increasing, quite similarly
to what found for the Abell sample.

[Despite this difference in the extension of the scaling regime, it is worth
noting that the value of $D_2 \sim 2.2$ obtained for both catalogs agrees
remarkably well with the slope $\gamma_2$ of the function $1+\xi(r)\propto
r^{\gamma_2}$ obtained by Calzetti, Giavalisco \& Meiksin (1992) for the
Abell clusters (note the $\gamma_2 \approx 3 - D_2$).]
Despite this difference in the extension of the scaling regime, it is
worth noting that the values of the $D_q$ dimensions are always quite
similar. For $q=2$ the resulting correlation dimension, $D_2 \sim 2.2$,
obtained for both catalogs agrees remarkably well with the slope $\gamma_2$
of the function $1+\xi(r)\propto r^{\gamma_2}$ obtained by Calzetti,
Giavalisco \& Meiksin (1992) for the Abell clusters (note the $\gamma_2
\approx 3 - D_2$).

Therefore, the overall emerging picture for the cluster distribution
indicates the presence of a well defined scaling behavior in the finite
scale range $15 \mincir r\mincir 60\hm$ for Abell sample, and more extended
to smaller scales for the ACO sample. This self--similar clustering is
followed by a breaking at larger scales. Despite this fact, we never succeed
to detect $D_q\simeq 3$, which would be the unique imprint of large--scale
homogeneity. However, this should not surprise, since, as previously shown
in \S3.1, the limited amount of statistics induces some {\em inertia} in
the large--scale detection of homogeneity, even for structures which are
intrinsically homogeneous at such scales. In any case, even adopting a
conservative point of view, we can reliably reject the hypothesis of a
purely fractal structure for the cluster distribution, extending up to
arbitrarily large scales.

The next question to be addressed in whether the different
behavior of Abell and ACO clusters is a spurious effect induced by limited
statistics and/or sample geometry, or they actually traces different
populations of cosmic structures.

To answer this question, we realized the same scaling analysis for
the set of simulated cluster samples presented in \S2.2. Figures 5 and 6
are the same as Figures 3 and 4, respectively, but for the artificial Abell
and ACO samples, so that they can be directly compared. The plotted data
are obtained by averaging over 50 different realization. In no cases
is there evidence of scale--invariant behavior extended on a scale--range as
large as that observed for the ACO sample. In the regime where the analysis
methods reliably work, the local dimension exhibits a smooth increasing
trend, similarly to the Abell case. Quite remarkably, the simulations,
which are intrinsically homogeneous at large scales, provide dimension
estimates at $r\magcir 50\hm$, which are quite similar to that displayed by
observational data. This further confirm that the observed cluster
distribution is consistent with a picture in which the clustering develops
induced by gravitational instability from nearly homogeneous initial
conditions. As for the method based on  excluding the self--point, it is
apparent in several cases the effect of overestimating the dimension at
small scales for $q=2$, which suggests that this method could have trouble
at small $q$ values, when dealing with such a limited statistics.

Therefore, the conclusion that we draw from the analysis of the synthetic
cluster samples is that the different scaling properties of Abell and ACO
clusters are likely to be real. While the Abell distribution can be
interpreted simply on the ground of a peak selection on a Zel'dovich
perturbed Gaussian background, the ACO statistics requires something more.
Conclusions go in the same direction as suggested by previous comparison
between simulated and observed cluster distributions. PVJ
found some significant differences between the clustering patterns of Abell
and ACO samples, the latter being systematically at variance with respect
to the clustering provided by the evolution of an initially Gaussian
fluctuation spectrum. A similar finding has also been found by Plionis,
Valdarnini \& Coles (1992) from the analysis of the topological genus. They
found a systematic ``meatball" shift for ACO clusters, which is
observed neither in the Abell sample nor in the same set of simulations.
The conservative assumption about the different statistics of the two
samples is that for ACO the higher sensitivity of the emulsion plates used
to construct the catalogue implies that it traces the cluster distribution
in a better way than Abell. Otherwise, one should be willing to accept the
idea of a real difference in the statistical properties of clusters in
different regions of the sky.

In Figure 7 we show the effects of including also clusters with $d>200\hm$
(left panel) and of not correcting for finite volume effects (right panel)
on the estimate of the correlation dimension $D_2$. In the first case, we
note some difference in the dimension estimate is introduced at small scales.
This shows that our procedure to correct for redshift selection works until
$P(z)$ takes very small values. In this regime, a larger number of objects
is needed to suppress noise effects. On the other hand, no significant
difference is induced by redshift selection at large scales. In this regime,
instead, the effects of border corrections are more important. As expected,
taking $f_i(r)=1$ in eq.(\ref{eq:ncor}) causes a spurious underestimate of
$D_2(r)$ for $r>40\hm$.

\section{Scaling from the $n$--nearest neighbor analysis}
\subsection{The method}
Although the method described in the previous section is in
principle valid for any real $q$, in practice the scaling
relation shown in eq.(\ref{eq:gp}) does not always hold  for
$q<2$, even for unambiguous multifractal sets. As we have already
said, low $q$ values emphasizes the  low density regions of the
point set. The scaling in this region is better followed by means
of the following scaling law,
\be
W(\tau,p)~=~{1\over N}\,\sum_{i=1}^N r_i(p)^{-\tau}~\sim
{}~p^{1-q}\,,
\label{eq:w}
\ee
where $r_i(p)$ is the radius of a sphere centered at point $i$,
which encompasses $n$ neighbors, and $p=n/N$. In an unlimited and
complete point set, $r_i(p)$ is just the distance from point $i$
to its $n$--th nearest neighbor. However, due to the sample
boundaries and the redshift and galactic latitude selection
functions of the surveys, the values of $r_i(p)$ have to be
corrected. Let us call $d_i^n$ the actual distance from cluster
$i$ to its $n$--th nearest neighbor within the survey. As we are
missing clusters at large distances and low latitudes, the
distance to the $n$--th nearest neighbor in a complete sample,
$r_i(p=n/N)$ should be less or equal to $d_i^n$. The correction
may be performed in the following way.

We compute $d_i^j$ for $j=1,...,n$. Then we consider the number
of neighbors, $n(d_i^j)$, within a ball of radius $d_i^j$, taking
into account the selection functions, but not the border
corrections, as in eq.(\ref{eq:ncor}). Note that $n(d_i^j)$ in
general is not an integer. The next step is to find the neighbors
$j_1$ and $j_2$ satisfying the inequality $n(d_i^{j_1})\le n \le
n(d_i^{j_2})$. Then the value $\tilde r_i(p=n/N)$ corrected for
selection functions is obtained just by the linear interpolation
\be
\tilde r_i(p)~=~d_i^{j_1}+(d_i^{j_2}-d_i^{j_1})\,{n-n(d_i^{j_1})\over
n(d_i^{j_2})-n(d_i^{j_1})}\,.
\label{eq:lin}
\ee
The radius has still to be corrected for the edge effects due to
the finite volume of the survey. This correction is performed
taking into account the fraction of the sphere of radius $\tilde
r_i(p)$, $f(\tilde r_i(p))$, falling within the boundaries of the
sample. The corrected radius is
\be
r_i(p)~=~\tilde r_i(p)\,\sqrt[3]{f(\tilde r_i(p))}\,.
\label{eq:rcor}
\ee
The corrected radius enters into eq.(\ref{eq:w}), in order to
obtain $q$ as a function of $\tau$.

The reliability of the $n$--nearest neighbor method to recover the
scaling properties of a fractal distribution has been discussed in details
by Borgani et al. (1993). After applying this method to the analysis of
several fractal distributions, it has been found that this algorithm
provides a considerable improvement in the estimate of the fractal dimension
in the $q<2$ regime.

\subsection{The results}
In Figs. 8 and 9 we show the scaling behavior of the function
$W(\tau,p)$ for $\tau=-2,-5,-10$.
The plotted errorbars are evaluated in the same way as for the
correlation integral method.

The analysis is just for the real cluster samples. The scaling is well
observed for both the Abell and ACO cluster samples. The oscillating
behavior of the dimension estimates observed in the plots have to be
interpreted as the fingerprint of the presence of lacunarity in the
catalogs. As $\tau$ decreases, the dimension increases showing
multifractality. The exact dimension values as well as the corresponding
statistical errors, are shown in Table 2. The scaling range for
$p=n/N$ in the linear regression fit has been $n=5 - 50$.
Figure 10 shows that this method is
less affected by the border correction than the correlation integral
method, since the dimensions shown on the top panels b and c are quite
similar. The correction for the selection functions explained in the
previous paragraph works well when the method is applied to the whole
catalog with depth 284 $\hm$. The dimension estimates obtained for the
closer and nearly complete sample (up to 200 $\hm$) agrees well with the
values obtained for the deepest and less complete sample (see
Figure 10).

As for the capacity dimension, we note that consistent values are
detected for Abell and ACO samples, with a value, $D_0\simeq 2.6$, which is
intermediate between that, $D_0\simeq 2.2$, detected for optical
galaxies (e.g., Mart\'{\i}nez et al. 1990;
Dom\'{\i}nguez--Tenreiro, G\'omez--Flechoso \& Mart\'{\i}nez 1993) and that,
$D_0\simeq 2.9$ estimated for QDOT--IRAS galaxies (Mart\'{\i}nez \& Coles
1994). Even
going to more and more negative $\tau$ values, no significant differences
are detected. This indicates that, apart from the different extension of
the scaling regime detected for $q\ge 2$, the two cluster samples display a
consistent clustering in both overdense and underdense regions.

\section{Discussion and conclusions}
We presented the results of a detailed scaling analysis for the spatial
distribution of Abell and ACO clusters. To this purpose, we resorted to
the fractal analysis approach, which has been shown to be a reliable tool
to characterize the statistics of the large--scale distribution of cosmic
structures (e.g., Mart\'{\i}nez et al. 1990; Valdarnini, Borgani \&
Provenzale 1992). The major problem in this kind of analysis is the rather
low amount of statistical information allowed by the available cluster
samples. For this reason, much care must be payed to disentangle the
effects of poor statistics in the interpretation of the the results.

To properly face these difficulties, we decided to apply two different
algorithms to estimate fractal dimensions, which are expected to work in
two different regimes; the correlation--integral method of
eq.(\ref{eq:gp}), which is expected to give reliable dimension estimates
within overdensities, and the $n$--nearest neighbor method of
eq.(\ref{eq:w}), which works
better inside the rarefied parts of the distribution. However, the limited
statistics does not represent the only problem. Further complications
arise due to the observational biases present in the cluster samples, such
as the finite size of the surveyed volumes and the selection effects both
in galactic latitude and in redshift. To account for these, we introduced
suitable corrections in the fractal dimension estimators and verified their
reliability on simulated cluster samples, with {\em a priori } known scaling
properties.

Based on the results presented in the previous sections, we can properly
answer the three questions addressed in the Introduction.
\begin{description}
\item[a)] The spatial cluster distribution displays a
scale--invariant behavior, which is however confined to a limited
scale range. The extension of this scaling regime is found to be
different for Abell and ACO clusters, respectively. The first
develops in the $15\mincir r\mincir 60\hm$ scale range, while the second
extends down to the smallest considered scales ($r\magcir 5\hm$). In
this scale range, the cluster distribution shows a marked
multifractal behavior. This is apparent from Figure 11, where we plot the
$D_q$ spectrum of multifractal dimensions, obtained from a log--log linear
regression over the scale--range ($15\mincir r\mincir 60\hm$), where both
Abell and ACO clusters develop the scaling behavior. For $q<2$ the
estimate is based on the $n$--nearest neighbor method, while for $q\ge 2$
we resort to the correlation--integral approach.

Note that in the former case the physical scale is not fixed a priori.
Instead, fixing the range of neighbor orders, larger and larger scales are
preferentially weighted in the $W(\tau, p)$ estimate as more and more
negative $q$ values are considered. Therefore, the decreasing trend of the
$D_q$ curve for $q<2$ could be not only due to multifractal scaling, but
also to the presence of scale--dependent clustering (see also Borgani et al.
1993). However, the rather smooth behavior of $D_q$ around $q=2$ indicates
that both analysis methods provide consistent dimension estimates inside
the overdensities.

For both samples the value of the correlation dimension is $D_2 \approx 2.2
\hm$ in agreement with the behavior of $1 + \xi(r)$ detected by
Calzetti et al. (1992).
Despite the different extension of the scaling regime inside the
overdensities, Abell and ACO clusters shows a remarkably consistent
multifractal spectrum in both $q>2$ and $q<2$ regimes.
\item[b)] The comparison with the cluster simulations based on the
Zel'dovich approximation shows that these substantially account for the
scaling of Abell clusters, while they do not reproduce the more extended
scaling regime displayed by the ACO sample. This suggests that the
distribution of ACO clusters requires something more than a peak selection
procedure on Zel'dovich perturbed Gaussian fluctuations, which were
otherwise able to generate the correct 2-- and 3--point cluster correlation
function (see also PVJ and Plionis, Valdarnini \& Coles 1992, for similar
conclusions about the different clustering of Abell and ACO redshift
samples).
\item[c)] The evidence for a limited scale--range of self--similar
clustering disproves the picture of a purely fractal structure of the
Universe, extending up to arbitrarily large scales (see also
Peebles (1993) for other indirect tests reaching to the same
conclusion). A different, but
related, question is whether we are able to detect the homogeneity scale
within the volume encompassed by the cluster samples. It is clear that, if
we define the homogeneity scale $L_h$ as that where $D_q(L_h)=3$, the
answer is no. Instead, at the largest considered scales ($\simeq 100\hm$)
we find $D_q\magcir 2$. However, we find the same result even analyzing
point distributions, which are intrinsically homogeneous at large--scales,
when using the same amount of statistics allowed by the cluster samples.
Therefore, our conclusion is that the spatial cluster distribution behaves
like a structure which becomes homogeneous at large scales around
$L_h\mincir 100\hm$.
\end{description}

\section*{Acknowledgements.} The authors wish to acknowledge the referee,
Joel Primack, for useful suggestions, which improved the
presentation of the results. SB acknowledges Queen Mary \& Westfield
College for its hospitality during the last phases of preparation of this
paper. We acknowledge the use of the QMW starlink and the IFIC computer
facilities where part of the calculations have been made. One of us (VM)
received partial support from a fellowship of the Conselleria d'Educaci\'o
i Ci\`encia de la Generalitat Valenciana. This work was partially supported
by the Direcci\'on General de Investigaci\'on Cient\'{\i}fica y T\'ecnica
(project number PB90--0416).

\newpage
\section*{\center References}
\vspace{0.2truecm}
\begin{trivlist}
\item[]Abell G.O., 1958, ApJ, 3, 211
\item[]Abell G.O., Corwin H.G., Olowin R.P., 1989, ApJS, 70, 1
\item[]Bahcall N.A., Soneira R.M., 1983, ApJ, 270, 20
\item[]Bahcall N.A., Cen R.Y., 1992, ApJ, 398, L81
\item[]Balian R., \& Schaeffer R., 1989, A\&A, 226, 373
\item[]Bardeen J.M., Bond J.R., Kaiser N., Szalay, A.S., 1986, ApJ, 304, 15
\item[]Blumenthal G.R., Dekel A., Primack J.R., 1988, ApJ, 326, 539
\item[]Borgani S., 1993, MNRAS, 260, 537
\item[]Borgani S., Coles P., Moscardini L., 1994, MNRAS, submitted
\item[]Borgani S., Coles P., Moscardini L., Plionis, M., 1994, MNRAS, 266,
524
\item[]Borgani S., Murante G., 1994, Phys. Rev. E, submitted
\item[]Borgani S., Murante G., Provenzale A., Valdarnini R., 1993b,
Phys. Rev. E, 47, 3879
\item[]Borgani S., Plionis M., Valdarnini R., 1993a, ApJ, 404, 21
\item[]Broadhurst T.J., Ellis R.S., Koo D.C., Szalay A.S., 1990, Nature,
343, 726
\item[]Calzetti D., Giavalisco M., Meiksin A., 1992, ApJ, 398, 429
\item[]Cappi A., \& Maurogordato S., 1992, A\&A, 259, 423
\item[]Coleman P.H., Pietronero L., 1992, Phys. Rep., 213, 31
\item[]Coleman P.H., Pietronero L., Sanders R.H., 1988, A\&A, 200, L32
\item[]Coles P., Melott A.L., Shandarin S.F., 1993, MNRAS, 260, 765
\item[]Croft R.A., Efstathiou G., 1994, MNRAS, submitted
\item[]Colombi S., Bouchet F.R., Schaeffer R., 1992, A\&A, 263, 1
\item[]Dalton G.B., Efstathiou G., Maddox S.J., Sutherland W.J.,
1992, ApJ, 390, L1
\item[]Davis M., Meiksin A., Strauss M.A., da Costa L.N., Yahil A., 1988,
ApJ, 333, L9
\item[]Davis M., Peebles P.J.E., 1983, ApJ, 267, 465
\item[]de Lapparent V., Geller M.J., Huchra J.P., 1986, ApJ, 302, L1
\item[]Dom\'{\i}nguez--Tenreiro R., G\'omez--Flechoso M.A.,
Mart\'{\i}nez V.J., 1993, ApJ, (in press)
\item[]Einasto J., Klypin A.A, Saar E., 1986, MNRAS, 219, 457
\item[]Geller M.J., Huchra J.P., 1989, Science, 246, 897
\item[]Grassberger P., 1988, Phys. Lett. A, 128, 369
\item[]Grassberger P., Procaccia I., 1983, Phys. Rew., A 28, 2591
\item[]Holtzmann J.A., Primack J.R., 1993, ApJ, 405, 428
\item[]Jones B.J.T., Coles P., \& Mart\'{\i}nez V.J., 1992, MNRAS, 259, 146
\item[]Kirshner R.P., Oemler A.Jr., Schechter P.L., Shechtman S.A.,
1981, ApJ, 248, L57
\item[]Kaiser N., 1984, ApJ, 284, L9
\item[]Klypin A.A., Kopilov A.I., 1983, Sov. Astr. Lett., 9, 41
\item[]Mann R.G., Heavens A.F., Peacock J.A., 1993, MNRAS, 263, 798
\item[]Mart\'{\i}nez V.J., Coles P., 1994, preprint
\item[]Mart\'{\i}nez V.J., Jones B.J.T., 1990, MNRAS, 242, 517
\item[]Mart\'{\i}nez V.J., Jones B.J.T., Dom\'{\i}nguez--Tenreiro R.,
van de Weygaert R., 1990, ApJ, 357, 50
\item[]Mart\'{\i}nez V.J., Portilla M., Jones B.J.T., Paredes S.,
1993, A\&A, 280, 5
\item[]Paladin G., Vulpiani A., 1984, Lett. Nuovo Cimento, 41, 82
\item[]Paladin G., Vulpiani A., 1987, Phys. Rep., 156, 147
\item[]Peebles P.J.E., 1980, The Large Scale Structure of the
Universe, Princeton University Press, Princeton
\item[]Peebles P.J.E., 1993, Principles of Physical Cosmology,
Princeton University Press, Princeton
\item[]Pietronero L., 1987, Physica A, 144, 257
\item[]Plionis M., Barrow J.D., Frenk C.S.F., 1991, MNRAS, 249, 622
\item[]Plionis M., Valdarnini R., 1991, MNRAS, 249, 46
\item[]Plionis M., Valdarnini R., Coles P., 1992, MNRAS, 258, 114
\item[]Plionis M., Valdarnini R., Jing Y.P., 1992, ApJ, 398, 12 (PVJ)
\item[]Postman M., Huchra J., Geller M., 1992, ApJ, 384, 404
\item[]Postman M., Spergel D.N., Sutin B., Juszkiewicz R., 1989, ApJ, 346, 588
\item[]R\'enyi A. 1970, Probability Theory (North Holland: Amsterdam)
\item[]Shandarin S.F., Zel'dovich Ya.B., 1989, Rev. Mod. Phys., 61, 185
\item[]Sutherland W., 1988, MNRAS, 234, 159
\item[]Sutherland W., Efsthatiou G., 1991, MNRAS, 258, 159
\item[]Valdarnini R., Borgani S., Provenzale, A., 1992, ApJ, 394, 422
\item[]Vogeley M.S., Park C., Geller M.J., Huchra, J.P., 1992, ApJ, 391, L5
\item[]White S.D.M., Frenk C.S., Davis M., Efstathiou G., 1987, ApJ, 313, 505
\item[]Yepes G., Dom\'{\i}nguez--Tenreiro R., Couchman H.M.P., 1992,
ApJ, 401, 40
\item[]Zel'dovich Ya.B., 1970, A\&A, 5,84
\end{trivlist}

\newpage
\section*{\center Figure Captions}
\vspace{0.2truecm}
\noindent
{\bf Figure 1.} The local dimension for the artificial Abell cluster sample,
extracted from a scale--dependent fractal structure, with homogeneity scale
$L_h\simeq 40\hm$ and $D=1$ for $r<L_h$ (see text). The results
are based on the correlation integral method. From top to bottom we plot
results for $q=2,3$ and 5. For each $r$ value, $D_q(r)$ is evaluated by
means of a 5--point log--log local linear regression on the slope of the
$Z(q,r)$ partition function. We compare results obtained when including the
center point in the evaluation of the probability $p_i(r)$ (circles), when
excluding it (squares) and when correcting for shot--noise effects (triangles).

\vspace{0.2truecm}
\noindent
{\bf Figure 2.} The same as in Figure 1, but for clusters extracted from a
purely fractal structure without large--scale homogeneity and dimension
$D=1$. Only results based on excluding the center points are shown.

\vspace{0.2truecm}
\noindent
{\bf Figure 3.} The $Z(q,r)$ partition function and the $D_q(r)$ local
dimension are plotted for the real Abell cluster sample. The analysis is
here realized by taking only clusters within $d<200\hm$. From left to
right we show results for $q=2,3$ and 5. Filled dots refer to the analysis
done by excluding the center points. The open dots for $q=3$ and 5 show the
local dimension obtained by correcting for shot--noise effects. Errorbars
for $Z(q,r)$ are evaluated as $1\sigma$ scatter between 100 bootstrap
resamplings. Errorbars in the local dimension are $3\sigma$ uncertainties
in the 5--point weighted least square fit.

\vspace{0.2truecm}
\noindent
{\bf Figure 4.} The same as in Figure 3, but for the ACO sample.

\vspace{0.2truecm}
\noindent
{\bf Figure 5.} The same as in Figure 3, but for the simulated Abell sample
(see text).

\vspace{0.2truecm}
\noindent
{\bf Figure 6.} The same as in Figure 3, but for the simulated ACO sample
(see text).

\vspace{0.2truecm}
\noindent
{\bf Figure 7.} Effects of corrections for boundaries and redshift selection
functions on the estimate of the correlation dimension $D_2$ for the Abell
sample. The left panel is for the whole sample, extending up to $d=284\hm$
and correcting for redshift selection. Central panel is the same as in
Figure 3. The right panel includes only clusters with $d<200\hm$ but
without correcting for boundary effects.

\vspace{0.2truecm}
\noindent
{\bf Figure 8.} The $W(\tau,p)$ partition function, along with the
corresponding local dimension, for the Abell sample. From left to right we
report results for $\tau=-2,-5$ and $-10$. Errors in $W(\tau,p)$ are
estimated from 100 bootstrap resamplings.

\vspace{0.2truecm}
\noindent
{\bf Figure 9.} The same as in Figure 8, but for the ACO sample.

\vspace{0.2truecm}
\noindent
{\bf Figure 10.} The same as in Figure 7, but for the $W(\tau,p)$ partition
function at $\tau=-2$.

\vspace{0.2truecm}
\noindent
{\bf Figure 11.} The multifractal dimension spectrum $D_q$ for Abell (solid
line) and ACO (dashed line) samples. For $q\mincir 2$, the estimate is
based on the $W(\tau,p)$ partition function, while for $q\magcir 2$ we
resort to the correlation integral method. In the first case the dimension
is evaluated over the range of $p =n/N$ values corresponding to $n= 5 -
50$, while in the second case it is evaluated over the scale range
$15\mincir r\mincir 60\hm$. Errorbars are standard deviations in the
weighted log--log linear regression of the corresponding partition
function.

\newpage
\begin{table}[tp]
\centering
\caption[]{$D_q$ dimensions for the Abell and ACO cluster samples
calculated by means of the correlation integral method. Also reported are
the corresponding standard deviations for the weighted linear regression.
The fit is realized over the scale--range $15\mincir r\mincir 60\hm$.}
\tabcolsep 7pt
\begin{tabular}{lccc} \\ \\ \hline
Sample & $D_2$ & $D_3$ & $D_5$ \\ \hline
Abell & $2.19\pm 0.05$ & $1.91\pm 0.04$ & $1.74\pm 0.03$ \\
ACO &   $2.21\pm 0.07$ & $1.95\pm 0.05$ & $1.80\pm 0.04$ \\ \hline
\end{tabular}
\end{table}

\newpage
\begin{table}[tp]
\centering
\caption[]{$D_q$ dimensions for the Abell and ACO cluster samples calculated
by means of the $n$--nearest neighbor method. For each value of $\tau$ the
obtained $q$ and $D_q$ is given, along with the corresponding statistical
errors. The estimate of the capacity dimension $D_0$
is shown in the first column. The fit is realized over the considered range
of $p$ values, with $p=n/N$, $n=5-50$.
of neighbor order.}
\tabcolsep 7pt
\begin{tabular}{lcccccccccc} \\ \\ \hline
 & & & \multicolumn{2}{c}{$\tau = -2$} & &\multicolumn{2}{c}{$\tau = -5$} & &
\multicolumn{2}{c}{$\tau = -10$} \\
Sample & $D_0$ & & $q$ & $D_q$ & & $q$ & $D_q$ & & $q$ & $D_q$ \\ \hline
Abell & $2.58\pm 0.04$ & & 0.21& $2.53\pm 0.04$ & & --0.80 &
$2.78\pm 0.06$ & & --2.15 & $3.17\pm 0.14$ \\
ACO & $2.65\pm 0.06$ & & 0.23& $2.59\pm 0.05$ & & --0.79 &
$2.80\pm 0.07$ & & --2.33 & $3.00\pm 0.10$ \\
\end{tabular}
\end{table}

\end{document}